\documentclass[acmsmall,screen]{acmart}
\usepackage{multirow}
\usepackage{makecell}
\usepackage[ruled,linesnumbered]{algorithm2e}
\usepackage{hyperref}
\AtBeginDocument{%
  }

\renewcommand\footnotetextcopyrightpermission[1]{}

\newcommand{\name}{ConZone}
\newcommand{\nameplus}{ConZone+}
\newcommand{\github}{https://github.com/DingcuiYu/ConZone}

\begin{document}

\title{\nameplus: Practical Zoned Flash Storage Emulation for Consumer Devices}

\thanks{This work is supported by the NSFC (62072177), Shanghai Science and Technology Project (22QA1403300) and the Open Project Program of Wuhan
 National Laboratory for Optoelectronics NO.2023WNLOKF004.}

\author{Dingcui Yu}
\email{dingcuiy@gmail.com}
\orcid{0009-0006-5344-2031}
\affiliation{%
  \institution{College of Computer Science and Technology, East China Normal University}
  \city{Shanghai}
  \country{China}
}

\author{Zonghuan Yan}
\email{yanhuan030824@163.com}
\orcid{0009-0006-5722-7371}
\affiliation{%
  \institution{College of Computer Science and Technology, East China Normal University}
  \city{Shanghai}
  \country{China}
  }

\author{Jialin Liu}
\email{52255901007@stu.ecnu.edu.cn}
\orcid{0009-0009-0669-2224}
\affiliation{%
  \institution{College of Computer Science and Technology, East China Normal University}
  \city{Shanghai}
  \country{China}
  }

\author{Yumiao Zhao}
\email{zhaoyumiao99@gmail.com}
\orcid{0000-0001-7803-2887}
\affiliation{%
  \institution{College of Computer Science and Technology, East China Normal University}
  \city{Shanghai}
  \country{China}
  }

\author{Yanyun Wang}
\email{51265901050@stu.ecnu.edu.cn}
\orcid{0009-0006-7963-7611}
\affiliation{%
  \institution{College of Computer Science and Technology, East China Normal University}
  \city{Shanghai}
  \country{China}
  }
  
\author{Xinghui Duan}
\email{danny@longsys.com}
\affiliation{%
  \institution{Longsys Electronics Co., Ltd}
  \city{Shanghai}
  \country{China}
  }

\author{Yina Lv}
\email{elainelv95@gmail.com}
\orcid{0000-0003-3971-3123}
\affiliation{%
  \institution{School of Informatics, Xiamen University}
  \city{Xiamen}
  \country{China}
  }
  
\author{Liang Shi}
\authornote{The corresponding author is Liang Shi.}
\email{shi.liang.hk@gmail.com}
\orcid{0000-0002-9977-529X}
\affiliation{%
  \institution{College of Computer Science and Technology, East China Normal University}
  \city{Shanghai}
  \country{China}
  }

\renewcommand{\shortauthors}{Dingcui et al.}

\begin{abstract}
Consumer-grade flash storage typically employs block interfaces for compatibility with file systems, but this results in significant mapping table overhead and write amplification penalties that degrade performance and endurance.
Zoned abstraction as an alternative, which organizes data into sequentially written zones, has been deployed in enterprise devices to solve these problems.
However, direct implementing zone abstraction in consumer devices leads to several challenges.
First, the limited volatile memory forces writes from multiple zones to compete for write buffers, whereas under the block interface, all writes can share a global write buffer.
This constraint limits the write performance of zoned storage and leads to severe write amplification. 
Second, some optimizations of the flash friendly file system (F2FS) tailored for block storage are incompatible with zone abstraction and result in substantial write amplification.
These challenges highlight the need for revisiting zoned storage design specifically for consumer devices.

To facilitate the understanding and efficient enhancement of software and hardware design for consumer-grade zoned flash storage, \name~is proposed as the first emulator designed to model the resource constraints and architectural features typical of such systems.
It incorporates essential components commonly deployed in consumer-grade devices, including limited logical to physical (L2P) mapping caches, constrained write buffers, and hybrid flash media management. 
However, \name~cannot be mounted with the file system due to the lack of in-place update capability, which is required by the metadata area of F2FS.
To improve the usability of the emulator, \nameplus~extends \name~with support for a block interface.
To ensure that the logical storage device with block access corresponds precisely to the metadata area of the file system, \nameplus~also provides a script that calculates the metadata size based on the capacity of the data area.
In addition, \nameplus~introduces several enhancements over the original version, including a configurable per-chip command queue, flexible block management, and compatibility with non-power-of-two block sizes. 
Users can explore the internal architecture and management strategies of consumer-grade zoned flash storage and integrate their optimizations with system software with \nameplus.
We validate the accuracy of \nameplus~by comparing a hardware architecture representative of consumer-grade zoned flash storage and comparing it with the state-of-the-art.
In addition, we conduct several case studies using \nameplus~to investigate the design of migrating and mapping mechanisms and explore the inadequacies of the current file system.
\end{abstract}




\maketitle

\section{Introduction} \label{introduction}
With the increasing demand for storage capacity across modern applications, high-density flash memory has been widely adopted in consumer-grade scenarios \cite{zms}\cite{luo2023cdb}\cite{luo2024cpf}. 
However, as flash density increases, the latency of read, program, and erase operations rises sharply, while flash memory endurance reduces significantly \cite{ISSCC-QLC}\cite{ISSCC-TLC}\cite{DTrim}\cite{lv2023mgc}. 
To ensure a satisfactory user experience, optimizing high-density flash-based storage systems has become essential \cite{luo2023cdb}\cite{luo2024cpf}\cite{song2023wear}\cite{gu2021dynamic}\cite{li2024elasticzram}\cite{li2023iosr}.

Consumer-grade flash storage (i.e., UFS, eMMC) is usually equipped with a block interface to support in-place updates and adapt to the granularity of host operations.
The management of block interface requires a logical-to-physical (L2P) mapping table, which is often stored in the on-device volatile memory (i.e., SRAM), and its capacity is one thousandth of the capacity of flash storage.
For example, a 256 GiB storage device requires 256 MiB of volatile memory for the L2P mapping table, significantly exceeding the available 1 MiB memory capacity \cite{zms}.
To reduce the memory capacity requirements for L2P mapping tables, the demand-based L2P caching  \cite{dftl}\cite{hotFTL}\cite{learnedftl} is designed to store most of the mapping information in flash memory and swapping it into the on-device volatile memory when needed.
Since the applications are diverse and accessed randomly, this naturally results in frequent cache misses and degraded read performance. 
In addition, the block interface lacks awareness of data validity and can only recognize host-side invalid data after receiving a TRIM command from the host system.
Several works have noted that the TRIM operation incurs a non-negligible execution delay \cite{itrim-2020}\cite{saxena2010flashvm}\cite{hyun2011trim}\cite{DTrim}.
As a result, file systems typically delay issuing TRIM commands by scheduling them to avoid interfering with foreground requests and batching multiple invalidation events into a single TRIM command\cite{itrim-2020}.
This delay in sending TRIM commands causes a mismatch between the host and the storage device: data already deleted or marked invalid by the file system may still be regarded as valid by the device for a certain period.
If the TRIM command is not timely, the flash controller may mistakenly treat stale data as valid and relocate it during garbage collection, leading to unnecessary writes that accelerate wear on high-density flash memory \cite{trim-advanced2017}\cite{itrim-2020}.

The zone interface \cite{zns} is developed to reduce memory overhead and offer enhanced flash endurance through coarse-grained zone mapping and host-managed data erasure.
The support for coarse-grained zone mapping is enabled by adopting sequential writes at the host level\cite{fms23zonedufs}\cite{zms}\cite{yan2024zufs}.
The host-managed data erasure avoids migrating invalidated data during garbage collection and thereby improves the endurance of flash memory.
At the host file system, mainstream consumer device manufacturers (e.g., Google, Samsung, and Huawei) have widely adopted the flash-friendly file system (F2FS) in their storage systems.
F2FS's native support for append-only writes naturally aligns with the sequential write constraints imposed by zoned storage architectures \cite{zms}\cite{wang2024eliminate}.
At the consumer device, zoned storage support is emerging.
The recent  JEDEC's Zoned UFS standard \cite{zonedstorageforUFS} enables zone abstraction in consumer-grade flash storage.
Furthermore, Samsung has explored specialized host-side I/O stack optimizations for zoned flash storage in mobile devices \cite{zms}.
However, there are some limitations of existing zoned flash storage in consumer devices.

Using zoned storage in consumer-grade storage requires 
optimizations not only for the storage firmware but also for the software stack.
The need for storage firmware optimization arises from the limited volatile cache and the usage of single-level cell (SLC) flash blocks, which are served as secondary write buffers\cite{shi2021understanding}\cite{lv2023access}\cite{li2022latency}\cite{zms}.
For write operations, the volatile write buffers are more likely to be flushed prematurely compared to block storage. 
In zoned storage, all open zones must share a small amount of write buffers (e.g., six open zones sharing two 384~KiB write buffers).
When the host writes to a different zone (e.g., writing cold data after hot data in F2FS) and the write buffers are already occupied, one must be flushed. 
In contrast, in block storage, all written data share a global write buffer, so changing the write address does not trigger a buffer flush.
If this flush is premature, meaning the amount of data is smaller than the programming unit, the data is temporarily redirected to the SLC-based secondary write buffer using partial programming.
This not only impacts write performance when migrating data from SLC to the high density flash blocks, but also causes write amplification and affects the endurance of the flash memory.
For read operations, the limited L2P cache capacity results in a higher probability of cache misses, necessitating frequent mapping table readings \cite{l2pmapping}\cite{learnedftl}. 
Although zone abstraction allows the use of coarser mapping granularity, premature flushes may lead to non-contiguous physical layouts even within a single zone. 
As a result, page-level mapping remains necessary for data written to the SLC region.
In summary, adopting zone abstraction in consumer-grade flash storage requires additional internal hardware design, including a careful redesign of volatile cache and SLC-based write buffers to avoid premature flushing, as well as a revised mapping table and L2P cache architecture to improve read efficiency and fully leverage the benefits of zone abstraction.

In addition to hardware-level constraints, the software stack, particularly the file system, also introduces compatibility issues with zone abstraction.
Although F2FS natively supports append-only writes and generally aligns with the sequential write model of zoned storage, many of its optimizations were originally designed for block storage and are therefore incompatible with zoned storage.
Specifically, F2FS incorporates optimizations for frequent small-grained synchronous in-place updates, which are common in consumer applications, by permitting data overwrites without modifying the corresponding node blocks, thereby reducing the volume of data written to flash memory \cite{zms}.
However, such behavior violates the sequential write constraints of zone abstraction.
As a result, small-grained synchronous in-place updates can generate more writes than using the block interface, degrading write performance and reducing flash endurance.
In addition, F2FS employs suboptimal garbage collection strategies for zoned storage.
It reserves an excessively large over-provisioned space and performs aggressive garbage collection even under light workloads\cite{patchfggcboost}\cite{patchgcgran}\cite{patchbggc}\cite{seo2023garbage}. 
Furthermore, shifting garbage collection to the host side extends the I/O path, reduces execution efficiency, and increases the likelihood of blocking user operations\cite{d2fs}\cite{zns+}.
These incompatibilities and inefficiency highlight the need for the optimization of the file system for consumer-grade zoned storage.

However, existing zoned namespace (ZNS) emulators do not incorporate key characteristics of consumer devices, such as constrained volatile write buffers and L2P caches, heterogeneous flash cells, and hybrid mapping schemes. 
This limitation makes it difficult to conduct firmware-level optimization research targeting consumer-grade zoned flash storage.
Moreover, commercially available ZNS solid state drives (SSDs) are primarily designed for enterprise scenarios, whose internal architectures differ significantly from those of consumer-grade devices, making them unsuitable for optimizing the I/O stack of consumer devices.
This underscores the pressing need for an emulation platform that can simulate zoned storage behavior in consumer scenarios.
To address these limitations, we propose \name, a dedicated emulator for consumer-grade zoned flash storage. 
\name~ models hardware features essential to consumer devices, including limited volatile memory for write buffers and L2P caches, hybrid address mapping mechanisms, and heterogeneous flash media management.
In addition, \name~reconstructs the internal processes of read, write, and erase operations to reflect the performance characteristics of actual zoned flash storage more accurately. 
This enables users to investigate the architectural design and internal management strategies of consumer-grade zoned flash storage.

To support system-level optimization evaluations, which require the support to perform in-place updates for F2FS metadata, we further propose \nameplus~as an extension of \name~by adding support for a block interface. 
The zone and block interfaces are exposed as two separate logical storage devices, enabling the use of multi-device formatting with \verb|mkfs.f2fs| to combine and format them together.
For accuracy, these two logically separate storage devices share the same physical flash storage.
\nameplus~ also provides a script to automatically calculate the metadata area size based on that of the data area, so that after formatting, the file system’s metadata region precisely maps to the block device address range.
Additionally, \nameplus~introduces several improvements over the original emulator, such as configurable per-chip command queue, flexible block management, and compatibility with non-power-of-two block sizes.
The code has also been refactored to improve modularity, making it easier to customize and extend for future research and development.
In the experiment, we validate the accuracy and features of \nameplus~by comparing an example of hardware configuration with the latest work that describes the organization and performance of zoned flash storage in consumer systems\cite{zms}.
We also present some case studies to demonstrate that \nameplus~can be helpful for further research on hardware design of consumer-grade zoned flash storage.
The source code of \nameplus~is publicly available at \href{\github}{\github}.
The contributions of this paper are as follows:
\begin{itemize}
\item We present \name, a lightweight framework for fast prototyping and internal analysis of consumer-grade zoned flash storage, enabling flexible exploration of firmware behaviors.
\item We introduce \nameplus, which adds support for formatting file systems for zoned flash prototypes and facilitates comprehensive analysis of the I/O stack from the file system to the device level on consumer-grade platforms.
\item We conduct case studies to explore firmware-level optimization strategies and reveal limitations in current file system designs when applied to zoned flash storage, offering guidance for future design.
\end{itemize}

The rest of the paper is organized as follows. 
Section \ref{sec:background_related} introduces the background and related work. 
Section \ref{sec:conzone} explains the internal structure of \name. 
Section \ref{sec:conzone+} introduces the internal structure of \nameplus.
Section \ref{sec:discussion} presents the limitation of the proposed design.
Section \ref{sec:experiment} discusses the evaluation results of \nameplus~by representing its flexibility and feasibility. 
Finally, Section \ref{sec:conclusion} gives the conclusion.

\section{Background and Related Work} \label{sec:background_related}
\subsection{Flash Storage in Consumer Devices}\label{subsec:bg_flash}
\begin{figure}[tp]
    \centering
    \includegraphics[width=0.99\linewidth]{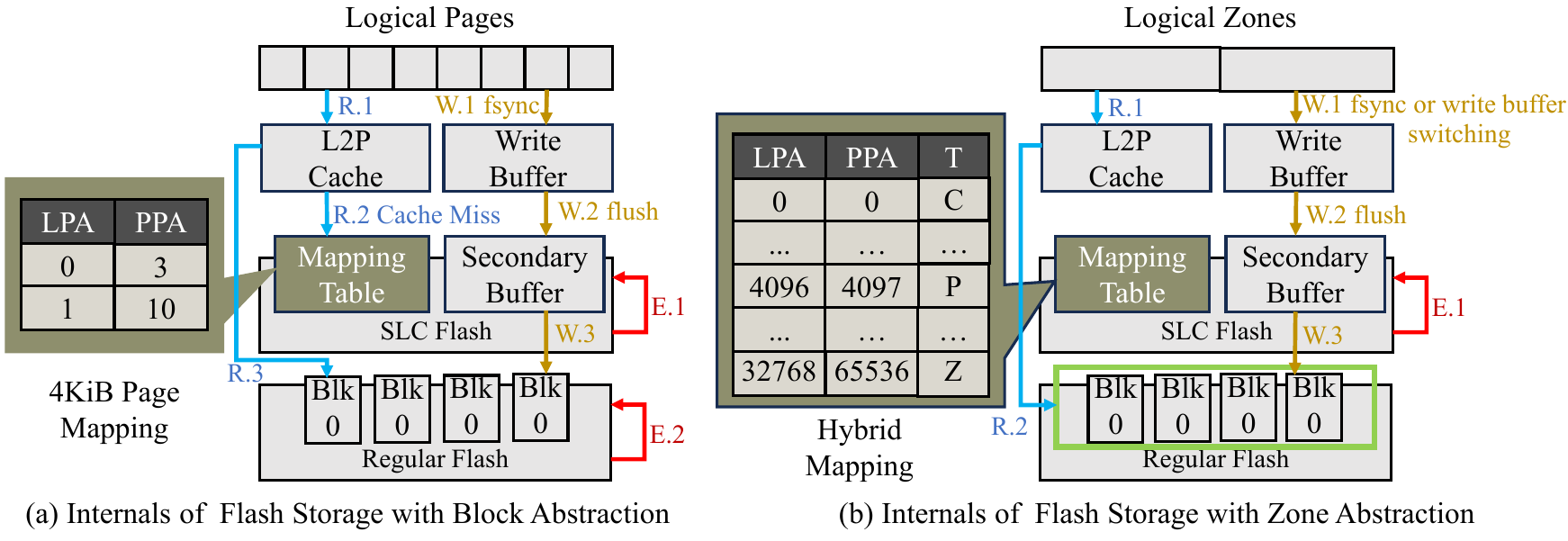}
    \caption{Comparison of legacy flash storage and zoned flash storage.}
    \Description{Diagram comparing legacy flash storage and zoned flash storage.}
    \label{fig:block_vs_zms}
    \vspace{-0.1in}
\end{figure}

Modern consumer-grade flash storage adopts a multi-channel, multi-way architecture to achieve higher performance. 
Data is striped across multiple channels and multiple ways.
A way is an independent group of NAND flash chips that can be accessed in parallel within a channel.
Fig. \ref{fig:block_vs_zms}(a) shows an example that four parallel flash blocks with the same offset across four chips (the "\verb|Blk 0|"s in Fig. \ref{fig:block_vs_zms}(a)), which collectively form a superblock. 
Data is striped over the four flash blocks. 
The stripe unit is composed of flash pages that are programmed simultaneously within each of the four blocks.
The size of each flash page is typically 16 KiB. 
With increasing flash density, the number of flash pages that must be programmed simultaneously also increases. 
For example, in triple-level cell (TLC) flash, the storage controller must program three pages at once. 
Therefore, in the Fig. \ref{fig:block_vs_zms}(a) example, the stripe unit size becomes 192 KiB (= 16 KiB × 3 × 4), and the programming granularity is 48 KiB (= 16 KiB × 3).

To maximize write throughput, the write buffer size is set with the stripe unit size. 
The number of write buffers corresponds to the number of open superblocks.
Unlike flash storage in servers, consumer-grade flash storage typically lacks power-loss protection\cite{zms}. 
When the memory page size is 4 KiB and the F2FS file system is used, the host issues data write requests in 4 KiB units\cite{patchf2fspagesize}.
When the host demands data persistence (e.g., via \verb|fsync|, as shown in Fig. \ref{fig:block_vs_zms}(a) W.1), the write buffer may not accumulate enough data to satisfy the programming granularity of high-density flash.
To address this, some flash blocks are programmed in SLC mode, forming pseudo-SLC (pSLC) blocks—still referred to as SLC blocks in this paper. 
As shown in Fig. \ref{fig:block_vs_zms}(a), these serve as secondary write buffers in the architecture.
For SLC flash, the storage controller allows partial programming, typically in 4 KiB units.
Consequently, data flushed prematurely from the write buffer can be written into the SLC buffer, satisfying the host’s persistence requirements (Fig. \ref{fig:block_vs_zms}(a) W.2).
Eventually, the accumulated data is written to the regular flash blocks (Fig. \ref{fig:block_vs_zms}(a) W.3).

Flash memory must be erased before it can be written to, and the erasure unit is a flash block.
When data is updated, the storage controller marks the original data as invalid and writes the updated data to a new physical location.
To manage this behavior, modern flash storage includes two key functional modules: the flash translation layer (FTL) and garbage collection (GC).
The FTL handles the dynamic mapping between logical addresses and physical addresses, ensuring that logical writes from the host can be redirected to appropriate physical locations as needed.
When data is accessed, the system must consult the L2P mapping to obtain the current physical address.
To accelerate reads, L2P entries are cached on demand in a volatile L2P cache \cite{dftl}\cite{hotFTL}.
During a read operation, the L2P cache is first checked to locate the target data (Fig. \ref{fig:block_vs_zms} (a) R.1).
If the required mapping does not present in the cache, the L2P entry must be retrieved from flash memory, resulting in degraded read performance (Fig. \ref{fig:block_vs_zms} (a) R.2).
Once the physical address is known, the storage controller reads the data accordingly (Fig. \ref{fig:block_vs_zms} (a) R.3).

GC addresses the mismatch between the fine write granularity and coarse erase granularity of flash memory.
It is responsible for relocating valid pages during block erasure and updating the corresponding L2P mappings(Fig. \ref{fig:block_vs_zms} (a) E.1, E.2).
To improve efficiency, GC is typically performed at the granularity of a superblock.
When performing GC in SLC, the storage controller can choose to migrate the valid data in the victim superblock either to the internal SLC space or to regular flash blocks.
This decision depends on the type of regular flash (e.g., TLC or QLC) and whether the user has enabled the option to buffer all writes in SLC.
For TLC, which employs single-step programming, users may choose to bypass the SLC and write data directly to regular flash blocks.
In this case, the storage controller migrates the valid data into the SLC space, as most data in SLC will naturally be relocated to regular flash blocks as the user continues to write.
For QLC, which uses two-step programming with a relatively long delay between the two steps, all user data must be buffered in SLC to prevent potential data loss\cite{2step-1}\cite{2step-2}\cite{2step-3}\cite{midas-touch}.
Therefore,  the storage controller migrates the valid data to regular flash blocks in this case to free up SLC space and prevent unnecessary long-term occupation.

\subsection{Zone Abstraction for Consumer Devices}\label{subsec:zms}
Under the zone abstraction, the host perceives flash storage as being divided into multiple zones, each of which must be written sequentially. 
Additionally, an erase operation (\verb|zone reset|) is introduced to reclaim logical space.
To support zone resets, the zone size must be aligned with the size of flash blocks.
Fig. \ref{fig:block_vs_zms} (b) illustrates an example where the zone size is equivalent to a superblock enclosed by green boxes to achieve higher per-zone performance\cite{ezns}.
Zone abstraction changes access patterns for flash storage.

First, it leads to more frequent premature flushes. 
F2FS can open up to six zones concurrently, with each open zone requiring a dedicated write buffer as each zone maps to a superblock. 
The write buffer capacity in consumer-grade flash storage is limited, so it is not feasible to allocate a dedicated write buffer for each open zone \cite{zms}.
As a result, when the system switches the active write zone, contention arises over limited write buffer space, causing data in other zones to be flushed prematurely (Fig. \ref{fig:block_vs_zms}(b) W.1).
Second, the sequential write constraint of the zone abstraction enables the use of coarser-grained mapping tables.
Since some data may be temporarily written to SLC flash blocks, the physical pages within a zone may not be physically contiguous, making hybrid mapping necessary.
The example in Fig. \ref{fig:block_vs_zms}(b) uses a three-level hybrid mapping of P (4 KiB page)–C (4 MiB chunk)–Z (zone), with a flag set in each page table entry to indicate the current mapping granularity.
The limited L2P cache can then store mappings that cover a wider logical address range, reducing the likelihood of L2P cache misses. 
Consequently, flash read is executed only once during the read process (Fig. \ref{fig:block_vs_zms}(b) R.1 and R.2).
Finally, zone abstraction shifts GC responsibility to the host. 
This eliminates the need for GC on regular flash blocks. 
However, GC is still required for the non-zone-managed SLC flash blocks to free up its space(Fig. \ref{fig:block_vs_zms}(b) E.1).

\begin{figure}[htp!]
    \centering
    \includegraphics[width=0.99\linewidth]{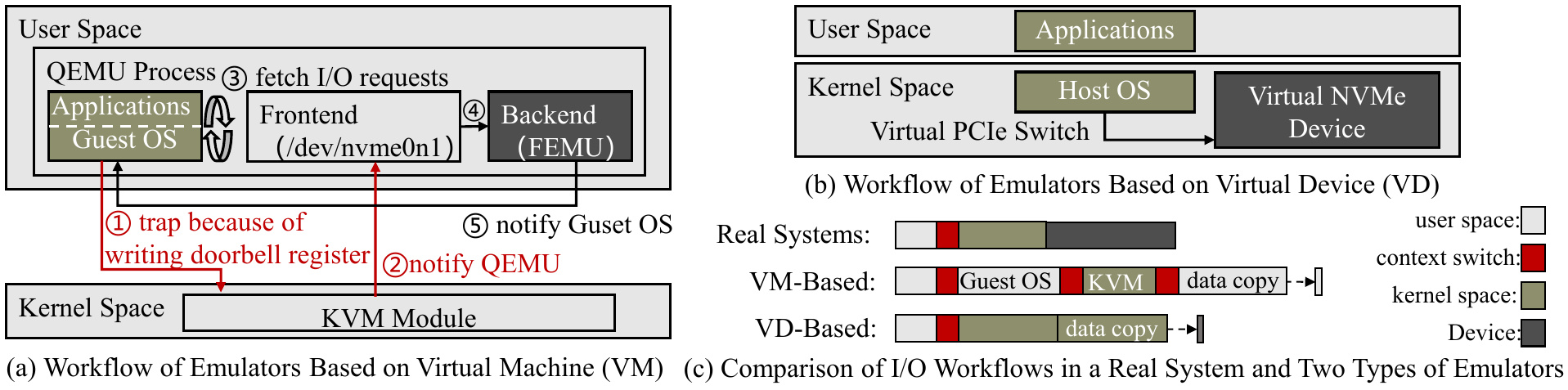}
    \caption{Comparison of current emulators.}
    \Description{Diagram comparing current emulators.}
    \label{fig:relatedwork}
    \vspace{-0.1in}
\end{figure}

\subsection{Existing Zoned Flash Storage Emulators}\label{subsec:existing}
Currently available zoned flash storage emulators are primarily designed for enterprise-grade ZNS SSDs, including FEMU \cite{femu}, ConfZNS \cite{confzns}, ConfZNS++\cite{confzns++} and NVMeVirt \cite{nvmevirt}.
Based on their implementation principles, these emulators can be categorized into two types: Virtual machine-based (VM-based) and virtual device-based (VD-based).
Their respective I/O workflows are illustrated in Fig. \ref{fig:relatedwork}(a) and (b).
In VM-based emulators (Fig. \ref{fig:relatedwork}(a)), applications run in the user space of the guest.
When an application issues an I/O request, it traps into the guest's kernel space.
The guest operating system then attempts to access a virtual NVMe device by writing to its doorbell register, which causes a VM-exit and traps into the kernel space of the host, where the KVM module takes over (\textcircled{1}).
KVM notifies the QEMU process, transferring control to the host's user space (\textcircled{2}).
Next, the QEMU frontend (e.g., a virtual \verb|/dev/nvme0n1| device in Fig. \ref{fig:relatedwork}(a)) reads from the guest kernel’s non-volatile memory express (NVMe) submission queue to retrieve the I/O request (\textcircled{3}), and then passes it to the QEMU backend (\textcircled{4}), which emulates device behavior in VM-based emulators.
For example, FEMU in Fig. \ref{fig:relatedwork}(a) is implemented as a QEMU backend.
The emulator typically copies data in memory and waits for a specified emulation delay before signaling completion to the guest kernel to mimic realistic device latency (\textcircled{5}).

\begin{table}[htp!]
   \centering
   \small
   \caption{Comparison of existing zoned flash storage emulators and \nameplus}
   \label{table:relate}
   \scalebox{0.9}{
	\begin{tabular}{|c|c|c|c|c|c|c|c|}
    \hline ~&~& \multicolumn{3}{|c|}{VM Based}& \multicolumn{3}{|c|}{VD Based}\\
    \hline ~& ~&\makecell{FEMU\\\cite{femu}}&\makecell{ConfZNS\\\cite{confzns}}&\makecell{ConfZNS++\\\cite{confzns++}}&\makecell{NVMeVirt\\\cite{nvmevirt}}&\makecell{\name\\\cite{conzone}}&\nameplus\\ 
    \hline 
    \multirow{7}*{Fidelity} & \makecell{Low-latency media\\  support}& No&  No&  No& Yes& Yes& Yes\\
    \cline {2-8}
    ~ & \makecell{Heterogeneous media\\ support} & No& No& No& No& Yes& Yes\\
    \cline{2-8}
    ~ &\makecell{ Limited write buffer\\ configuration} & Yes& No& No& No& Yes& Yes\\
    \cline{2-8}
    ~ & \makecell{L2P cache \\configuration} & No& No& No& No & Yes& Yes\\
    \cline{2-8}
    ~ & L2P mapping & Linear& Linear& Zone & Linear & Hybrid& Hybrid\\
    \cline{2-8}
    ~ & \makecell{Garbdage collection \\support} & No& No& No& No& Yes& Yes\\
    \cline{2-8}
    ~ & Per-chip command queue & No& No& No& No& No& Yes\\
    \hline 
    \multirow{4}*{Versatility}  & \makecell{\# of emulated devices\\ at the same time}& many& many& many& 1& 1& 1\\
    \cline{2-8} 
    ~ & \# of namespace& 1& 1& 1& 2& 2& 2\\
    \cline{2-8} 
    ~&\# of SSD instance& 1& 1& 1& 2& 2& 1\\
    \cline{2-8}
    ~&\makecell{Namespace \\to\\ SSD instance\\ mapping} & 1-to-1& 1-to-1& 1-to-1& 1-to-1& 1-to-1& 2-to-1\\
    \hline
	\end{tabular}
    }
\end{table}

In contrast, VD-based emulators implement the flash device as a kernel module.
The virtual device is connected via a virtual PCIe switch to the PCIe root complex, allowing the host system to recognize it as a native PCIe device (Fig. \ref{fig:relatedwork}(b)).
VD-based emulators can simulate device latency more precisely, which is especially critical for low-latency media.
As shown in Fig. \ref{fig:relatedwork}(c), dashed arrows represent emulation delays determined by user-specified device latencies and the flash device states.
VM-based emulators introduce additional context switches before device emulation begins. 
These extra context switch delays are unpredictable and cannot be offset through simple timing adjustments.
FEMU identifies this issue and attempts to mitigate it by commenting out a kernel call to avoid trapping into the host kernel \cite{femu}. 
Instead, it lets the QEMU process poll the guest's NVMe submission queue to fetch new I/O requests promptly.
However, this workaround requires a deep understanding of the NVMe driver in the kernel, significantly increasing the complexity and usability barrier for users.
Given that consumer-grade flash storage often uses SLC caches, whose access latency is on the order of tens of microseconds, and that user-friendliness is a key concern, we chose to build our consumer-grade zoned flash storage emulator based on the VD-based architecture.

We also evaluated how existing emulators support the specific characteristics of consumer-grade zoned flash storage, as shown in Table \ref{table:relate}.
Unfortunately, none of them fully meet our requirements.
Specifically, the in-development mainline version of FEMU supports write buffers but lacks L2P cache and a fully functional FTL in ZNS mode\cite{femu}.
ConfZNS, built upon an earlier released version of FEMU, implements diverse static zone mapping and an accurate I/O timing model \cite{confzns}.
ConfZNS++ builds upon ConfZNS by adding dynamic zone mapping and quantifying the latency of I/O management operations \cite{confzns++}.
Both ConfZNS and ConfZNS++ lack the write buffer and L2P caching support.
NVMeVirt was the first emulator to support virtual NVMe devices, and its ZNS support largely follows FEMU’s design.
Its ZNS mode does not support heterogeneous flash cells, L2P caching, or hybrid address mapping.
To address these limitations, we developed \name, which emulates the essential internal hardware components required for consumer-grade zoned flash storage.

\section{\name~Internals}\label{sec:conzone}

\subsection{Overview}\label{overview}
Although the interface standard for consumer-grade flash storage is UFS, we choose to develop \name~on NVMeVirt\cite{nvmevirt}, a simulation platform designed for NVMe flash storage, based on the following considerations. 
First, UFS relies on the MIPI M-PHY interface, which is specifically tailored for SoCs and low-power embedded systems. 
Typical PC hosts do not provide native support for MIPI M-PHY, and dedicated mobile development boards that support UFS are costly and difficult to access. 
In contrast, the NVMe ecosystem is mature, widely supported across general-purpose hosts, and has a lower development barrier.
Second, while UFS and NVMe differ mainly in terms of bandwidth and I/O concurrency (i.e., UFS uses MIPI M-PHY with only two data lanes and supports up to 32 SQ/CQ queue pairs, whereas NVMe uses PCIe, supports multiple lanes, and up to 65,536 queue pairs), these differences can be largely approximated by tuning configurable parameters in our platform.
Power consumption optimization is also a major focus of UFS, but is beyond the scope of our study.
Finally, despite differences at the interface level, UFS and NVMe storage devices share highly similar abstractions at the storage controller level, including essential components such as FTL and GC mechanisms, which are the focus of our research. 

\begin{figure}[htp!]
    \centering
    \includegraphics[width=0.99\linewidth]{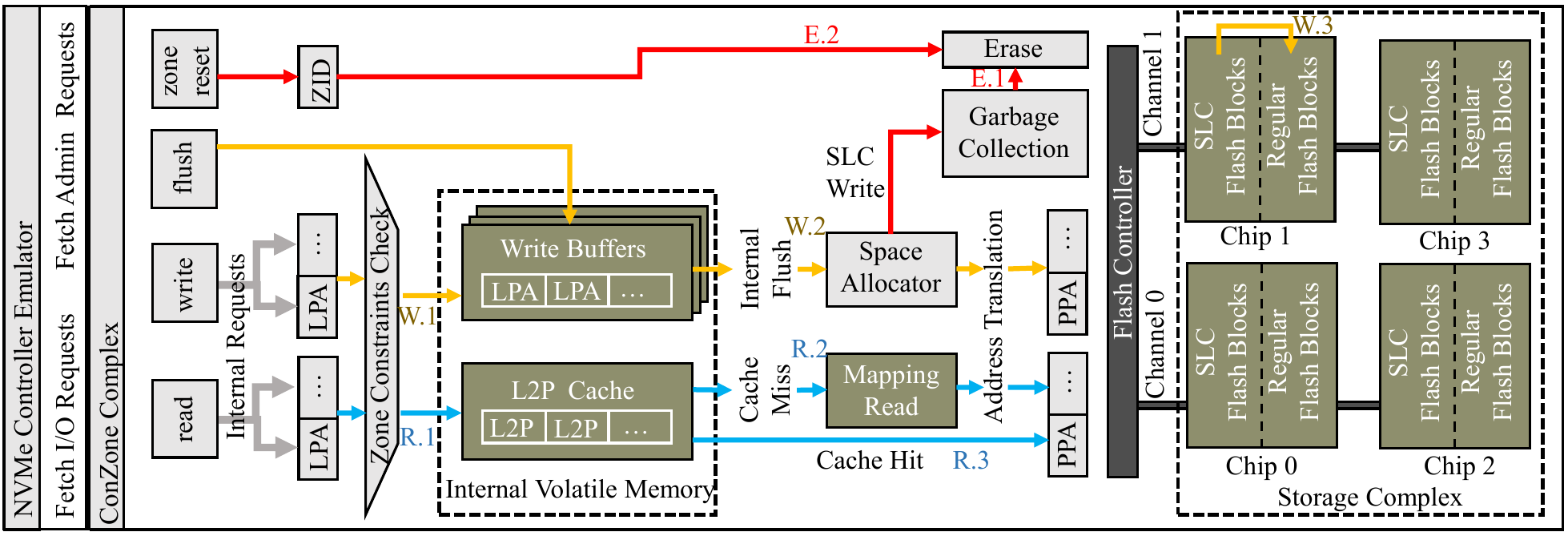}
    \caption{Internals of \name.}
    \label{fig:overview}
    \vspace{-0.1in}
\end{figure}

Fig. \ref{fig:overview} illustrates the key components of \name.
\name~implements the essential hardware modules required to emulate consumer-grade zoned flash storage during read, write, and erase (i.e., \verb|zone reset|) operations.
These modules include limited volatile memory for L2P cache and write buffers, a mapping table fetching mechanism for handling L2P cache misses, a space allocator for dynamic hybrid mapping, a garbage collection module for reclaiming space in SLC flash, and support for heterogeneous flash media.
The timing model is located at each parallel unit (i.e., channels and chips) of the storage complex.

For write operations, \name~limits the number of available write buffers and maintains a mapping between write buffers and zones (Fig. \ref{fig:overview} W.1).
When the active write zone changes, data in the corresponding write buffer is either flushed to regular flash blocks or temporarily redirected to SLC flash blocks (Fig. \ref{fig:overview} W.2).
The space allocator dynamically assigns physical addresses based on the write location.
Then, considering the status of the parallel units and media latency, \name~calculates the latency and ends the write simulation of the current request.
Once enough data has accumulated, the data temporarily stored in the SLC is migrated to regular flash blocks (Fig. \ref{fig:overview} W.3).
Additionally, when the host issues a flush command, all data in the write buffers is immediately flushed to flash.

For read operations, \name~uses a flat one-level mapping table to locate all data.
Page-level mapping is adopted at first.
As the zone fills up, \name~gradually increases the mapping granularity.
Specifically, once the physically contiguous data reaches a chunk (4 MiB) or a full zone, fine-grained page mappings are merged into a single coarser-grained entry and cached in the L2P table.
We refer to this mechanism as hybrid mapping.
Each read request of the host first queries the L2P cache (Fig. \ref{fig:overview} R.1).
If an L2P cache miss occurs, the hybrid mapping mechanism may require multiple flash reads to fetch the corresponding L2P mapping entry, potentially causing performance fluctuations (Fig. \ref{fig:overview} R.2).
After retrieving the physical address, \name~begins the actual data read (Fig. \ref{fig:overview} R.3), and the L2P cache is subsequently updated.
Similar to writes, the final read latency is determined based on the parallel unit status and after that the read simulation is completed.

For erase operations, \name~embeds a complete garbage collection mechanism to reclaim SLC flash blocks (Fig. \ref{fig:overview} E.1).
This process includes victim block selection, valid page migration, block erasure, and mapping table updates.
In addition, \name~supports two types of SLC garbage collection: in-place garbage collection and migration-based collection, where data in SLC flash blocks is relocated to regular flash blocks.
For regular flash blocks, space reclamation is entirely host-controlled. 
When the host issues a zone reset command, the system directly erases the corresponding regular flash blocks and updates the mapping table accordingly (Fig. \ref{fig:overview} E.2).

\subsection{Write Path: Hybrid Media and Limited Write Buffer}\label{zmswrite}
\begin{figure}[tp]
    \centering
    \includegraphics{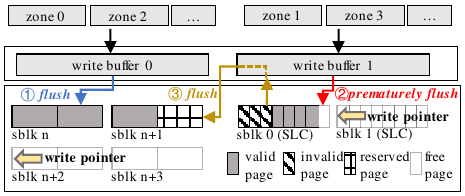}
    \caption{Write Path of \name}
    \label{fig:writepath}
    \vspace{-0.1in}
\end{figure}

Fig.\ref{fig:writepath} illustrates the write path of \name.
Writes from different zones are first directed to their corresponding write buffers for temporal aggregation.
The flush path is determined by both the volume of accumulated data and the physical data layout within the current zone.
If the data volume is sufficient to meet the programming granularity, the data is directly flushed to a regular flash block (\textcircled{1}).
Otherwise, it is temporarily flushed to an SLC flash block (\textcircled{2}).
Once the data stored in the SLC flash block, combined with newly arriving data, reaches the programming granularity, the previously written data in the SLC flash block is fetched and invalidated (represented by striped blocks in Fig.\ref{fig:writepath}), and the combined data is flushed to a regular flash block(\textcircled{3}).

The space allocator of \name~uses write pointers to manage physical address assignment.
Specifically, a write pointer is bound to a free superblock and advances within the superblock according to predefined rules.
Once the end of a superblock is reached, the write pointer is reassigned to a new free superblock.
By repeatedly advancing the write pointer, \name~can reserve a contiguous region of physical addresses.
The granularity of space allocation is controlled based on the amount of data to be reserved.
In SLC flash blocks, space is allocated at the page level.
In contrast, for regular flash blocks, space is allocated at the zone level (depicted as square-patterned blocks in Fig.~\ref{fig:writepath}) to ensure that data belonging to the same zone remains physically contiguous within the regular flash.
When writing to such a reserved region, the physical address can be directly computed from the logical offset within the zone.

\textbf{Conflicting Zone-Write Buffer Mapping:}
Given the limited hardware resources in consumer-grade storage devices, it is impractical to dedicate a separate write buffer for each open zone.
To accommodate this constraint, users are allowed to configure both the total number of available write buffers and the size of each write buffer.
Currently, \name~supports two mapping strategies between zones and write buffers.
The first is a fully-associative mapping, in which any zone can be dynamically bound to any write buffer.
The second is a modulo-based mapping, where the target write buffer is determined by taking the modulus of the zone ID with the total number of write buffers.
In the fully-associative mode, when all write buffers are occupied, the system selects the buffer with the largest amount of accumulated data for flushing, and reassigns it to serve incoming writes from a new zone.
By default, \name~uses the fully-associative mapping strategy, though users may also define custom mapping rules as needed.

\textbf{Heterogeneous Media and Extended Timing Model:}
To maximize the parallelism of flash storage, the write pointer moves to the next flash block after programming one unit, cycling through all blocks within a superblock before proceeding to the next programming unit of the first flash block.
As a result, the iteration behavior of the write pointer varies according to the programming granularity of the underlying media type.
In addition, since SLC flash blocks in consumer-grade flash storage are typically converted from regular flash blocks, their effective capacity is correspondingly reduced.
In addition, we extend the timing model in \name~to account for the heterogeneity of flash memory technologies.
Table \ref{table-timing} illustrates the default access latencies for different types of flash cells.
These values are primarily derived from published academic studies, while the read latency for SLC blocks is based on discussions with industry engineers.
Users could configure the first $n$ flash blocks of each chip to be treated as SLC blocks and specify their corresponding access latencies.

\begin{table}[!htp]
\vspace{-0.1in}
   \centering
   \caption{Latency for Different Media in \name }
   \label{table-timing}
   \scalebox{1}{
	\begin{tabular}{|c | c |c |c|}
    \hline & SLC&TLC&QLC\\ 
    \hline Program& 75~\textmu s\cite{slc-timing}&  937.5~\textmu s\cite{tlc-write-timing}& 6400~\textmu s\cite{qlc-timing}\\
    \hline Read& 20~\textmu s& 32~\textmu s\cite{tlc-write-timing}& 85~\textmu s\cite{qlc-timing}\\
    \hline
	\end{tabular}
 }
 \vspace{-0.1in}
\end{table}

\subsection{Read Path: Hybrid Mapping and L2P Cache Management}
\begin{figure}[tp]
    \centering
    \includegraphics{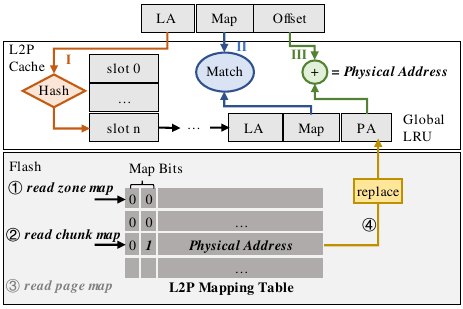}
    \caption{Read Path of \name}
    \label{fig:readpath}
    \vspace{-0.1in}
\end{figure}

Fig.\ref{fig:readpath} illustrates the read path in \name.
Upon receiving a read request, the system first queries the L2P cache.
\name~sequentially translates the original logical page address into a logical zone address (LZA), a logical chunk address (LCA), and finally the logical page address (LPA), and attempts to match each level in the L2P cache in order (\uppercase\expandafter{\romannumeral1}).
If a match is found (\uppercase\expandafter{\romannumeral2}), the physical address (\uppercase\expandafter{\romannumeral3}) is computed using the offset between the original logical page address and the matched logical address.
If no match is found in the cache, the corresponding L2P mapping entry must be retrieved from flash memory.
Similarly, the mapping table is queried hierarchically using the LZA, LCA, and LPA (\textcircled{1}, \textcircled{2}, \textcircled{3}).
\name~leverages two reserved bits in each mapping table entry to indicate three mapping granularities.
If the reserved bits of the corresponding mapping entry are set (\textcircled{2}), a new L2P cache entry is created and inserted into the L2P cache (\textcircled{4}).
When the cache is full, entries are evicted following the least-recently-used (LRU) policy.

\begin{figure}[!htp]
    \centering
    \includegraphics{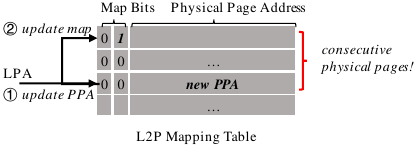}
    \caption{\name's hybrid mapping mechanism, LPA for logical page addresses and PPA for physical page addresses}
    \label{fig:hybridmap}
    \vspace{-0.1in}
\end{figure}

\textbf{Hybrid Mapping: }\label{hybridmap}
\name~supports the aggregation of mapping table entries with contiguous logical and physical addresses into a single L2P cache entry to enhance read performance.
The FTL adopts a flat, one-level mapping scheme to maintain the logical-to-physical address translation.
As illustrated in Fig.\ref{fig:hybridmap}, when data is flushed from the write buffer, the corresponding logical-to-physical mapping is established and the mapping table is updated accordingly (\textcircled{1}).
Since a series of contiguous regular flash blocks is reserved for each zone, \name~can efficiently determine whether a group of mapping entries can be aggregated by checking whether the physical addresses align with chunk or zone boundaries (\textcircled{2}).
In contrast, data temporarily written to SLC flash blocks is not eligible for aggregation, as continuity in physical addresses cannot be guaranteed.

\textbf{Management of L2P Cache: }\label{l2pcache}
L2P cache entries contain three fields: the logical address, the mapping granularity, and the corresponding physical address.
To accelerate cache lookups, logical addresses are distributed across multiple hash buckets.
\name~traverses the cache entries within the relevant hash bucket in a hierarchical manner in the order of the logical zone address (LZA), logical chunk address (LCA), and logical page address (LPA).
A cache hit occurs when both the logical address and its mapping granularity match.
In the case of an L2P cache miss, the corresponding mapping entry must be retrieved from flash memory.
This introduces a key challenge: \textbf{\textit{how to determine the aggregation level of the current zone before accessing the mapping table}}.
One possible solution is to maintain a bitmap that tracks the mapping granularity for all logical addresses, incurring a capacity overhead of approximately 0.006\%.
For a 1 TiB flash storage, this would require about 64 MiB of volatile memory, which is impractical for consumer-grade devices.
An alternative approach is to perform multiple reads like the hierarchical lookup used in the L2P cache.
Specifically, the system first assumes that the address is mapped at the zone level, fetches the corresponding LZA mapping entry, and checks its mapping bits.
If the check fails, the system proceeds to fetch the LCA-level entry, and so on.
However, this approach causes L2P cache misses to incur multiple flash reads, leading to degraded read performance.
We conduct a case study in Section \ref{case} to evaluate the performance impact of such multiple fetches.

\subsection{Erase Path: Composite Garbage Collection}
SLC flash blocks and regular flash blocks operate under distinct management models.
The validity and invalidity of SLC blocks are entirely managed by the storage controller, whereas regular flash blocks are fully managed by the host.
Based on this distinction, \name~employs separate GC mechanisms for the two types of flash blocks.
For SLC flash blocks, \name~implements a full GC process.
It first selects a victim superblock based on the number of valid pages, migrates the valid data to another location, erases the victim superblock, and then returns it to the SLC free superblock pool.
\name~can choose to migrate the valid data in the victim superblock either to the internal SLC space or to regular flash blocks as described in Section~\ref{subsec:bg_flash}.
For regular flash blocks, \name~adopts a simplified GC process.
When the host resets a zone, \name~directly erases the regular flash blocks previously allocated to that zone and invalidates corresponding data in SLC flash blocks.
\name~also updates mapping table entries of that zone to maintain consistency.

\section{\nameplus~Internals}\label{sec:conzone+}
\textbf{Motivation: }
Since standalone zoned storage devices cannot be formatted with the F2FS file system, we developed \nameplus~to support a broader range of experimental scenarios, such as evaluating file system performance or benchmarking SQLite. 
To enable this, we required a flash storage exposing a block interface.
NVMeVirt comes with a prototype implementation based on the Samsung 970 Pro SSD\cite{nvmevirt}.
However, directly employing this prototype introduces several limitations.
First, the prototype lacks key consumer-grade flash features, such as an L2P cache and hybrid media management.
Second, NVMeVirt initializes two independent \verb|SSD| instances, each maintaining its own latency model, thus failing to simulate resource contention accurately.
ZMS\cite{zms}, a recent state-of-the-art study on zoned storage firmware and system design for mobile devices, divides a single flash storage device into two Logical Units (LUs): one exposing a block interface and the other a zoned interface.
This LU concept is analogous to the namespace abstraction in the NVMe protocol, where a storage device is partitioned into logically independent units that still share the same physical resources.
Therefore, to reuse consumer-grade flash components implemented in \name~and accurately simulate contention between two namespaces accessing the same physical media, we extend \name~into \nameplus~.

\begin{figure}[tp]
    \centering
    \includegraphics[width=0.99\linewidth]{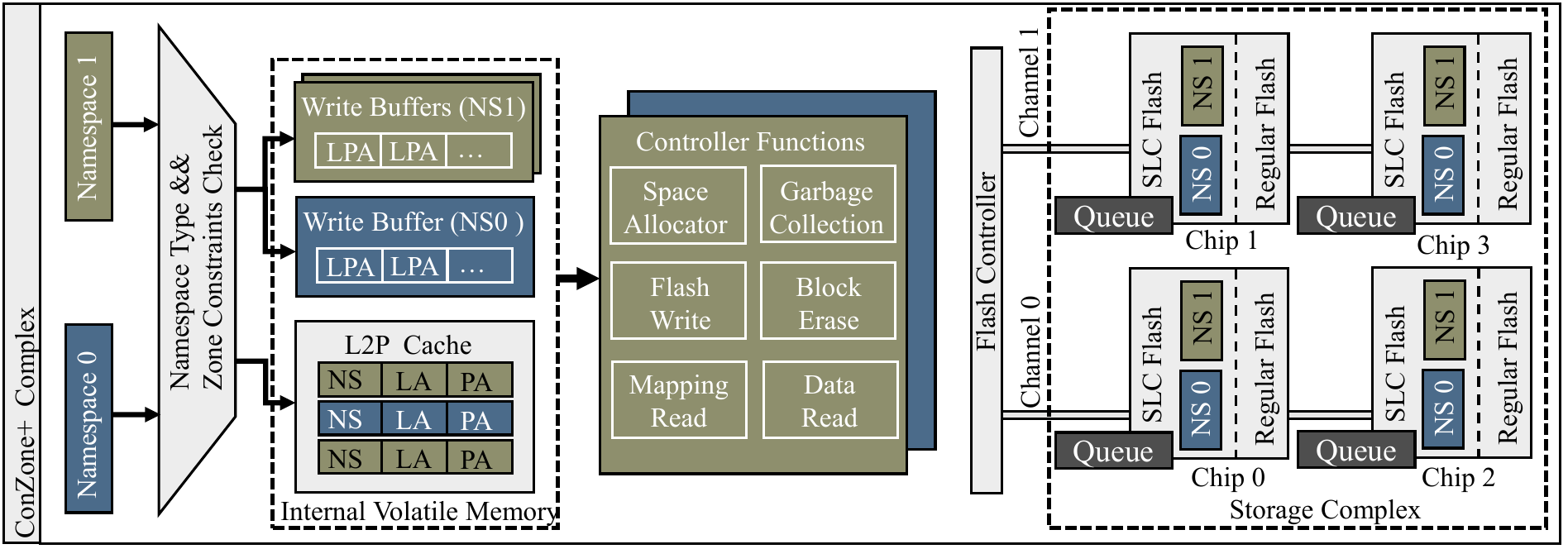}
    \caption{Internals of \nameplus.}
    \label{fig:conzoneplus}
    \vspace{-0.1in}
\end{figure}

\textbf{Overview: }
\nameplus~ modifies NVMeVirt’s initialization procedure to support multiple namespaces sharing a single \verb|SSD| instance, and integrates block interface support into the existing codebase.
The internal structure of \nameplus~ is illustrated in Fig.~\ref{fig:conzoneplus}.
Specifically, upon receiving a new request, \nameplus~ first checks the namespace type. 
If the request targets a zoned interface, it performs corresponding validations accordingly.
For write operations, each namespace is allocated a dedicated portion of the write buffer to ensure isolation.
For read operations, since the logical address spaces of the two namespaces are independent, the L2P cache is extended with an additional namespace identifier (denoted as “NS” in Fig.~\ref{fig:conzoneplus}) alongside the logical address (LA) to distinguish mapping entries from different namespaces.
Correspondingly, the two namespaces are each assigned an independent controller instance, responsible for their respective L2P mapping management, space allocation, and garbage collection.
However, once data is written to the shared \verb|SSD| instance, both namespaces inevitably contend for the same physical resources, leading to potential interference.
By default, \nameplus~ designates namespace 0 as the block interface, intended to store file system metadata.
To minimize latency and improve reliability for such critical data, and based on discussions with industrial flash storage experts, \nameplus~ places all data from namespace 0 into SLC flash blocks.
Consequently, within the storage complex, the SLC flash pool is partitioned into two segments, each assigned to a different namespace.

Additionally, \nameplus~ enables users to customize the request queue management policy for each parallel unit\cite{liu2022fair}.
For example, to coordinate host and internal maintenance requests in a manner that maximizes overall responsiveness.
At the mapping layer, \nameplus~ also supports fine-grained block-level management within each superblock, providing enhanced flexibility for future zone resource management\cite{ezns}.
These extensions offer greater versatility for exploring and optimizing consumer-grade zoned storage architectures.
The specific differences can be found in Table~\ref{table:relate}.

\subsection{Support for File System Metadata}
Algorithm~\ref{alg:ns-init} presents the pseudocode for our namespace initialization logic. 
When the variable \verb|BASE_SSD| is set to \verb|CONZONE_PROTOTYPE|, the emulation platform introduced in this work is enabled.
Users can configure the number and properties of namespaces by defining the following variables in \verb|ssd_config.h|: \verb|NR_Namespace| specifies the total number of namespaces; \verb|NS_TYPE[]| defines the type of each namespace; \verb|LOGICAL_NS_SIZE[]| and \verb|PHYSICAL_NS_SIZE[]| specify the logical and physical size of each namespace, respectively.
During initialization, the system iterates over all namespaces (Line1).
For each namespace, if the prototype is  \verb|CONZONE_PROTOTYPE|(Line2), an FTL instance is allocated (Line3), and static parameters unrelated to SSD instances, such as garbage collection thresholds, are configured. 
Subsequently, the logical and physical sizes of the namespace are set accordingly (Lines4--5).
For namespaces of other types, the system follows the default NVMeVirt initialization routine (Lines~7--9), which includes independent SSD and FTL instance setup.
After the loop completes, if the prototype is \verb|CONZONE_PROTOTYPE| (Line12), a shared SSD instance is initialized, followed by the instantiation of FTL instances for each namespace with respect to the shared physical media (Lines13--15).
With this, the initialization of \nameplus~ is completed.
Correspondingly, during simulation termination, the shared SSD instance is released first, followed by the release of each namespace's FTL instance.

\begin{algorithm}[t]
\caption{Namespace Initialization in NVMeVirt with \nameplus~ Extensions}
\label{alg:ns-init}
\KwIn{\texttt{NR\_Namespace}, \texttt{NS\_TYPE[\,]}, \texttt{LOGICAL\_NS\_SIZE[\,]}, \texttt{PHYSICAL\_NS\_SIZE[\,]}}

\For{$i \leftarrow 0$ \KwTo $\texttt{NR\_Namespace} - 1$}{
    \If{\texttt{BASE\_SSD} == \texttt{CONZONE\_PROTOTYPE}}{
        Allocate  FTL instance for namespace $i$\;
        Set logical size $\leftarrow$ \texttt{LOGICAL\_NS\_SIZE}[i]\;
        Set physical size $\leftarrow$ \texttt{PHYSICAL\_NS\_SIZE}[i]\;
    }
    \Else{
        Initialize standalone SSD instance\;
        Initialize FTL for namespace $i$\;
    }
}

\uIf{\texttt{BASE\_SSD} == \texttt{CONZONE\_PROTOTYPE}}{
    Initialize shared SSD instance\;
    \For{$i \leftarrow 0$ \KwTo $\texttt{NR\_Namespace} - 1$}{
        Instantiate assigned FTL instance for namespace $i$\;
    }
}
\end{algorithm}

\textbf{Namespace Size Configuration: }
When loading \nameplus~, a total physical capacity must be specified, which includes the physical sizes of both namespaces.
For the namespace used to store user data, the logical size is user-defined.
The SLC flash blocks serving as a secondary write buffer are transparent to the user.
Since zoned-interface namespaces do not require over-provisioned (OP) space, the total physical size of this namespace equals the logical size plus the capacity of the blocks programmed as SLC.
For the namespace used to store file system metadata, the logical size requires additional consideration.
This is because \verb|mkfs.f2fs|, when formatting multiple devices, concatenates them into a single logical address space and calculates metadata requirements based on the total capacity.
In other words, the starting address of the data section in F2FS may not align with the starting offset of the namespace that is used to store data.
If misaligned, some user data may be written to the metadata device, leading to unintended interference in experiments.
To address this, we modified the source code of \verb|mkfs.f2fs| to output the layout information.
Users can try different logical sizes for the metadata namespace and inspect the output to verify correctness.
This process is not time-consuming, since the metadata section in F2FS is aligned with the zone size.
The zone size of F2FS matches that of the underlying zoned storage.
Typically, one zone is sufficient to store metadata for small-capacity devices (e.g., 4 GiB).
F2FS reserves the first zone and starts placing metadata from the second zone.
Therefore, in practice, we recommend starting with a reservation of two zones and increasing the size incrementally as needed.
We include the modified version of \verb|mkfs.f2fs| with additional output in the \nameplus~ source tree for ease of use.
The physical size of the metadata namespace should be aligned with the flash superblock size and account for necessary OP space, based on the configured logical size.
Finally, the total size of both namespaces must be summed to form the load-time configuration for \nameplus~.
To further simplify this process, we also provide an interactive configuration script in the source tree of the \nameplus~ repository.

\subsection{Support for Per-Chip Command Queue}
To further improve the responsiveness of consumer-grade flash storage to user requests, request scheduling is necessary for commands submitted to the device \cite{liu2022fair}.
\nameplus~introduces this support and implements a basic mechanism where user requests can preempt background (non-user) operations.
Specifically, \nameplus~adds a command queue to each parallel unit, where requests are normally processed in a FIFO manner.
To prevent SLC-to-regular flash block migration from impacting user requests, \nameplus~suspends the migration process and gives priority to user I/O.
Before the migration completes, the mapping table is not updated, and the space allocator does not allocate the migrating superblock for new writes.
Therefore, such preemption does not introduce any consistency issues.
Suspending background requests can reduce the blocking time experienced by the user.
However, when migrations occur frequently, the average response latency remains difficult to improve.
Addressing this challenge calls for a more holistic design of the storage architecture, including adjustments to the size of the SLC-based secondary write buffer (see Section~\ref{sec:discussion}), which is beyond the scope of this work.

\subsection{Support for Flexible Block Management}
In the original \name, both space allocation and reclamation (i.e., erasure) are performed at the granularity of a superblock.
This limits the flexibility of configuring small zones.
If a small zone is defined, its erasure cannot be performed immediately but must wait for other zones within the same superblock to be reclaimed together.
To address this limitation, \nameplus~extends support for sub-blocks within a superblock, where each sub-block corresponds to a regular flash block.
Users can choose whether to enable sub-block mode depending on their experimental needs.

\subsection{Compatible with Non-Power-of-Two Block Sizes}
The zone abstraction defined by the NVMe standard currently does not allow non-power-of-two zone sizes.
Therefore, when the underlying flash media is TLC, setting up a compliant zone becomes infeasible.
\name~provides a temporary workaround by aligning the zone size and redirecting the overflow beyond TLC flash block limits to SLC flash pages. 
In contrast, \name~addresses this limitation by using the zone capacity field, which does not have the power-of-two restriction, and modifying the space allocator to define zone boundaries based on zone capacity instead of zone size.
\name~still aligns the physical space with the aligned power-of-two zone size, resulting in a portion of unused space. 
However, this does not affect the accuracy of emulation. 
Since the regular flash blocks managed via the zone interface do not require garbage collection, the free space is not treated as any thresholds.
Meanwhile, NVMe developers are actively working to support non-power-of-two zone sizes, and we believe this limitation will eventually be lifted~\cite{no-po2-zonesize}.

\section{Limitations and Discussions}\label{sec:discussion}
\textbf{Persistence of L2P Mapping Table Updates: }
Since the L2P mapping table needs to be persisted into flash memory, how to update the mapping table entries is a design challenge.
Currently L2P log is used within consumer-grade flash storage to accumulate L2P mapping table updates.
The L2P log is flushed to flash memory when several updates are accumulated, and the flushing back of the L2P log may block host requests.
In addition, how to seek the address of the L2P mapping entries after flushing back, and whether other structures of page tables need to be used are also topics that need to be explored.
This feature will be realized in future work.

\textbf{Dynamic Flash Block Conversion: }
For high-density flash memory, the number of blocks configured as SLC presents a trade-off between capacity loss and block migration overhead.
Because SLC programming reduces the storage density per cell, the capacity of an SLC-configured flash block is smaller than that of a regular block, leading to a reduction in the overall usable capacity.
On the other hand, a too-small SLC region results in frequent data migration, which increases write request latency and the write amplification ratio.
Since all data must be written to the SLC region first (see Section~\ref{subsec:bg_flash}), data in the SLC flash blocks must be migrated to regular flash blocks once free SLC blocks become insufficient.
Dynamic adjustment of SLC capacity has been widely studied in the literature\cite{convert-1}\cite{convert-2}\cite{convert-3}\cite{shi2021understanding}.
To facilitate future research, we plan to integrate block type conversion functionality into our future work.

\section{Evaluation}\label{sec:experiment}
In this section, we evaluate the accuracy and functionality of \nameplus, and present several case studies. 
The following aspects are primarily explored:
\begin{enumerate}
\item How precisely can \nameplus~emulate the performance of zoned flash storage for consumer devices?
\item What are the benefits and challenges in zoned flash storage for consumer devices?
\item How does F2FS behave differently when running on zoned storage compared to conventional block storage?
\item How does the design of zoned flash storage internals affect I/O performance?
\end{enumerate}

\subsection{Evaluation Setup}
\textbf{Evaluation Environment: }
We implemented \nameplus~on an HP Z8 G4 workstation equipped with two Intel Xeon Silver 4114 2.20 GHz processors and 94 GiB of memory, running Linux kernel version 6.12.16. 
The implementation consists of approximately 2,500 lines of code.
Due to the lack of consumer devices equipped with zoned flash storage, we extensively referenced and compared information disclosed in recent academic work~\cite{zms}, which we refer to as "ZMS" throughout this paper. 
The test results obtained from our zoned storage emulation platform are labeled as "ConZone - Zoned Device".
As the internal structure of flash storage in existing consumer devices is not publicly disclosed, we additionally implemented a block flash storage prototype based on the descriptions provided in~\cite{zms}. 
This prototype is referred to as "ConZone - Block Device" in our experiments.
To validate the accuracy of our block storage emulation, we also conducted the same tests on a Google Pixel 6 smartphone for comparison.
In addition, we compare our design with the latest versions of FEMU\cite{femu} and NVMeVirt\cite{nvmevirt} both running under the same hardware configuration. 
To enable formatting with the F2FS file system, we added an additional storage device using a block interface in FEMU. 
For NVMeVirt, we configured an additional namespace that supports the block interface. 
Notably, the \verb|SSD| instance associated with the block-interface namespace is independent from the one used for the zone-interface namespace.
Their corresponding results are denoted as "FEMU - ZNS" and "NVMeVirt - ZNS", respectively.
We use a flexible I/O tester (FIO)\cite{fio} and mobibench\cite{mobibench} to generate synthetic workloads for benchmarking.

\textbf{Configuration: }
To mitigate the impact of virtualization on emulator performance, we reserved two dedicated CPU cores for executing \nameplus. 
We configured \nameplus's parameters by referencing ZMS. 
Specifically, we set the flash type to TLC and allocated the number of SLC flash blocks according to the requirements for write aggregation and alignment. 
Additionally, we configured two parallel channels, each connected to two chips. 
After consultation with industry engineers, we set the channel bandwidth to 3200~MiB/s, referencing the standard bandwidth of UFS~4.0 and accounting for redundant read data overhead.
We limited the queue depth to a maximum of 32. 
The zone size was set equal to the superblock size to maximize sequential read and write performance. 
For write operations, we configured the programming unit to 96~KiB to emulate simultaneous writes to two planes within a single chip. 
All zones shared two write buffers, each 384~KiB in size, consistent with the configuration used in ZMS. 
For read operations, we set the L2P cache size to 1~MiB. 
The total storage capacity of the emulated device was set to 4~GiB.

\begin{figure}[tp]
    \centering
    \includegraphics[width=0.9\linewidth]{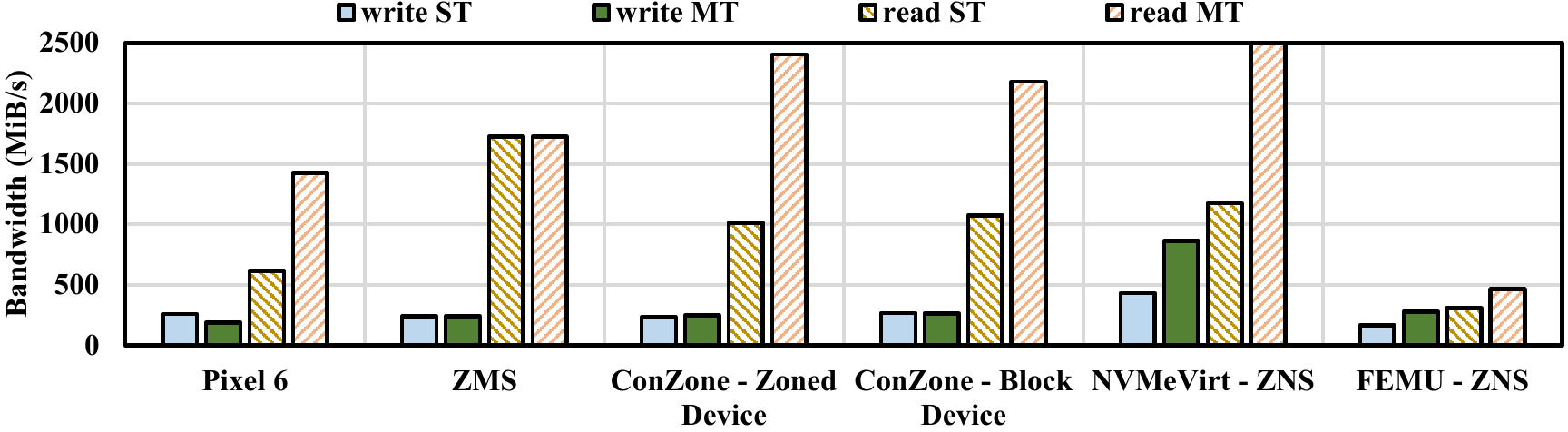}
    \caption{Comparison of Sequential I/Os with Different Platforms. ST denotes single thread and MT denotes 4 threads}
    \label{fig: accuracy}
    \vspace{-0.1in}
\end{figure}

\subsection{The Accuracy of \nameplus}
We use FIO to perform 512~KiB sequential read and write operations, following the experimental setup of ZMS. 
Fig~\ref{fig: accuracy} shows the sequential I/O bandwidth results, where ST denotes single-threaded execution and MT denotes multi-threaded execution using four threads.
For write performance, both in ST and MT cases, the results of ConZone - Zoned Device and ZMS, as well as those of ConZone - Block Device and Pixel 6, are closely aligned, indicating the accuracy of \nameplus's write emulation.
In terms of read performance, ConZone - Zoned Device and ZMS exhibit different trends. 
In ZMS, the number of threads has little impact on sequential read performance, whereas in ConZone - Zoned Device, the read bandwidth in the MT case is nearly twice that of the ST case.
There are two possible reasons for this. 
First, since flash read latency is on the order of tens of microseconds, read performance is influenced not only by the storage but also by kernel behavior and CPU performance. 
The CPU used in the ZMS (SM8350) may have better single core performance than our CPU (e.g., SM8350 has a higher Geekbench 6 single-core score than Xeon 4114).
In contrast, our CPU has stronger multi-core performance. 
When FIO runs in MT mode, it utilizes multiple cores to issue requests, thereby compensating for the performance gap in ST execution.
Interestingly, the read performance trend observed on the Pixel~6 mirrors that of both ConZone - Zoned Device and ConZone - Block Device, with the bandwidth roughly doubling as the thread count increases. 
This consistency suggests that our platform is capable of capturing real-world read performance trends to a certain extent.
Second, ZMS does not disclose the details of its L2P cache implementation. 
Differences in the L2P cache between \nameplus~and ZMS may also contribute to the observed variation in read bandwidth.
The write performance of NVMeVirt - ZNS exceeds that of ZMS.
There are two reasons.
First, the ZNS prototype in NVMeVirt does not consider size constraints on the write buffer. 
Notice that the request size configured in FIO is 512 KiB. 
If the write buffer were limited to 384~KiB, as in ZMS, the emulator would continuously reject writes due to insufficient write buffer space. 
To make the test feasible, we set the write buffer size for each zone to 512~KiB.
Second, NVMeVirt does not simulate metadata and data I/O contention because it uses separate \verb|SSD| instances for different namespaces, eliminating potential conflicts.
In sequential read scenarios, where the L2P cache hit rate is typically high, cache capacity has minimal impact on read performance. 
Therefore, the read bandwidth of NVMeVirt - ZNS is similar to that of ConZone - Zoned Device.
The overall performance of FEMU - ZNS is relatively low, primarily because it runs inside a virtual machine. 
The additional context switching overhead along the I/O path affects the fidelity of its simulation (see Section~\ref{subsec:existing}).

\subsection{The Benefits and Challenges of Zoned Flash Storage}
\begin{figure}[tp]
    \centering
    \includegraphics[width=0.9\linewidth]{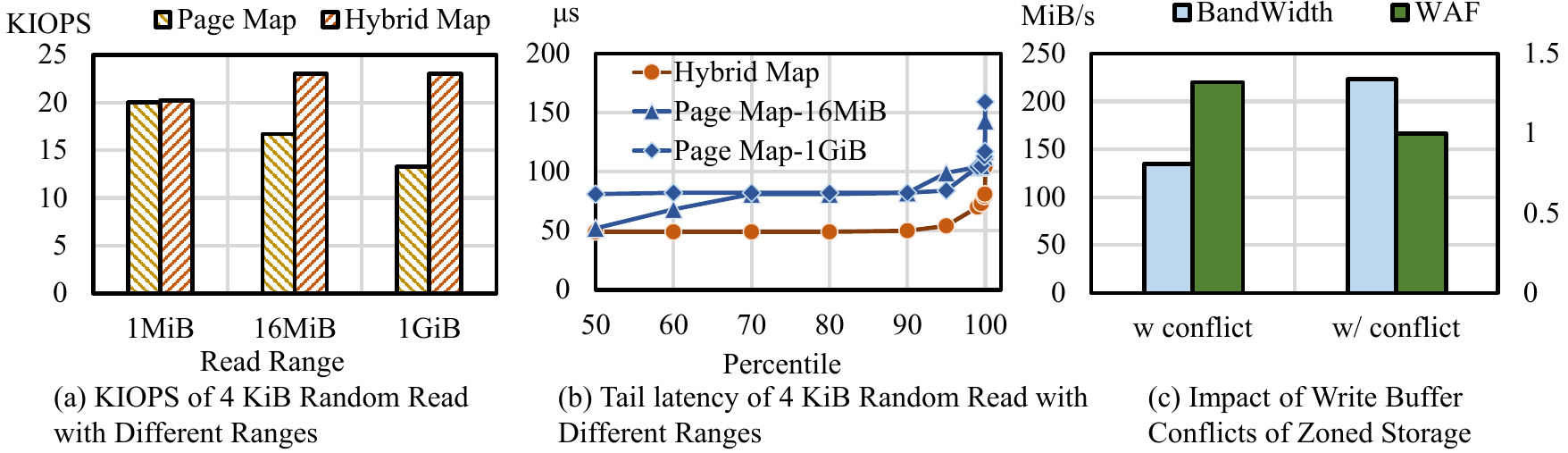}
    \vspace{-0.1in}
    \caption{Benefits and Challenges of Zoned Storage}
    \label{fig: challenges_benefits}
    \vspace{-0.1in}
\end{figure}

\textbf{Benefits of Hybrid Mapping: }  
Compared to block interfaces, zone interfaces can adopt coarse-grained mapping tables, thereby reducing the demand on the L2P cache and offering significant advantages in read performance.  
In this section, we evaluate the impact of page mapping and hybrid mapping on 4~KiB random read performance with zoned flash storage.  
All tests are conducted under the same data volume but with varying read ranges.
As shown in Fig.~\ref{fig: challenges_benefits}(a), when the read range is 1~MiB, the KIOPS of both mapping mechanisms reaches 20.2K, as all mapping entries can be accommodated in the L2P cache for both page and hybrid mappings.  
However, when the read range increases to 16~MiB and 1~GiB, the KIOPS of page mapping drops by 16.5\% and 33.5\%, respectively, compared to the 1~MiB baseline.  
This degradation is caused by the rising L2P miss rates, which increase to 41.1\% and 98.1\% under page mapping.
The benefits of hybrid mapping are also evident in terms of tail latency.  
As shown in Fig.~\ref{fig: challenges_benefits}(b), the tail latency of random reads under hybrid mapping remains consistently around 50~\textmu s, since all relevant mapping entries can still reside in the cache.

\textbf{Challenges of Write Buffer Conflicts: }  
As described in Section~\ref{introduction}, write buffer conflicts can occur in zoned storage systems.  
Fig.~\ref{fig: challenges_benefits}(c) illustrates the negative impact of such conflicts.
We design the following test methodology to evaluate this effect.  
First, odd-numbered and even-numbered zones are assigned to two separate write buffers.  
Then, two threads are used to write one zone's worth of data each, with a write granularity of 48~KiB to intentionally trigger premature flushing of the write buffers.  
When the zones being written to share the same parity (i.e., both odd or both even), write buffer conflicts occur; otherwise, no conflicts arise.
The results demonstrate a 65\% increase in write bandwidth when buffer conflicts are eliminated. Furthermore, the write amplification factor (WAF), defined as the ratio of device writes to host writes and a key metric for flash longevity, is reduced by 24\%.
These findings highlight the importance of avoiding write buffer conflicts in the design of zoned flash storage systems.

\begin{figure}[tp]
    \centering
    \includegraphics[width=0.9\linewidth]{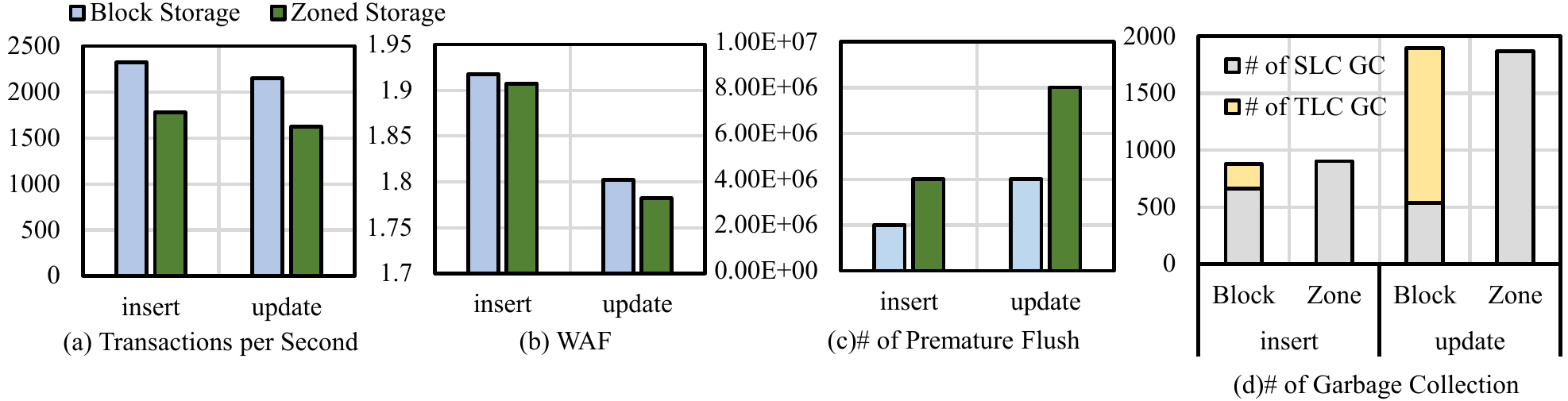}
    \caption{Impact of Zoned Storage on SQLite Transaction Performance and Write Amplification}
    \label{fig: sqlite}
    \vspace{-0.1in}
\end{figure}

\textbf{Challenges of Zoned Storage in SQLite Transaction Processing:}
Beyond internal design challenges such as write buffer contention, zoned storage also grapples with issues arising from file system limitations, particularly its lack of support for in-place updates.
This problem is especially pronounced in SQLite database transaction processing, which frequently invokes \verb|fsync|. 
We utilized Mobibench to generate 2,000,000 SQLite insert and update operations, with SQLite configured in WAL mode. 
The experimental results are presented in Fig. \ref{fig: sqlite}.
As shown in Fig. \ref{fig: sqlite}(a), the Transactions per Second (TPS) under block device storage is 30.5\% and 32.2\% higher than that of zoned storage during insert and update operations, respectively. 
This is attributed to the file system writing more node data and more frequently triggering file garbage collection when using zoned storage. 
These factors lead to an increase in read/write data volume and, consequently, impact TPS.
Conversely, Fig. \ref{fig: sqlite}(b) indicates that the in-device WAF of zoned storage is slightly lower than that of block storage. 
This is because, in SLC flash blocks, both storage types primarily need to erase invalid blocks left after data is naturally relocated by the host. 
In contrast, block storage necessitates garbage collection on TLC flash blocks, as illustrated in Fig. \ref{fig: sqlite}(d).
Furthermore, Fig. \ref{fig: sqlite}(c) reveals more frequent premature write buffer flushes in zoned storage. 
This further points to the presence of write buffer contention shown in this section.

\subsection{Case Study: F2FS Behavior on Zoned Storage vs. Block Storage}
To quantify the differences in F2FS file system behavior when facing zoned storage versus block storage, we conduct a case study.
During mounting, F2FS supports various mount options, including \texttt{adaptive} and \texttt{lfs}. 
The \texttt{adaptive} option allows for self-adaptive selection of whether to perform in-place updates, while \texttt{lfs} permits only out-of-place updates. 
When using block storage, the default mount option is \texttt{adaptive}. 
However, with zoned storage, only the \texttt{lfs} mount option is supported.
Furthermore, F2FS allows users to configure the frequency of GC triggers via \verb|sysfs|. 
Given that zoned storage does not inherently perform garbage collection, F2FS executes a significantly more aggressive GC on zoned storage.
In this experiment, we compare the following configurations:
\begin{itemize}
    \item \textbf{block-adaptive}: The storage interface type is block, with the mount option set to \texttt{adaptive}.
    \item \textbf{block-lfs}: The storage interface type is block, with the mount option set to \texttt{lfs}.
    \item \textbf{zoned}: The storage interface type is zone, utilizing the default GC policy in F2FS.
    \item \textbf{zoned-config}: The storage interface type is zone, but the F2FS GC frequency is adjusted to match that of the block interface. This includes configuring \verb|gc_max_sleep_time|, \\ 
    \verb|gc_min_sleep_time|, and \verb|gc_no_gc_sleep_time|.
    Additionally, enhancements specific to zoned storage were disabled by setting both \verb|gc_boost_zoned_gc_percent| and \\ \verb|gc_no_zoned_gc_percent| to 0, thereby reducing the trigger frequency of foreground GC\cite{patchfggcboost}\cite{patchbggc}.
\end{itemize}
We format a storage device with a data area size of 4 GiB as F2FS. 
In the first step, we use FIO to issue 4 KiB direct I/O writes, sequentially writing a 2 GiB large file (approximately 60\% capacity utilization) to the disk.
Subsequently, in the second step, we perform 4 KiB random updates on this file, updating a total of 8 GiB of data.
During the first step of writing, we observe that the write bandwidth of zoned storage (average 86 MiB/s) is lower than that of block storage (average 106 MiB/s). 
This is because, even when FIO requests direct I/O, F2FS on zoned storage still writes data to the page cache first before synchronizing it to the device. 
This behavior is necessary because direct I/O alone cannot guarantee sequential writes on zoned storage\cite{patchdio}\cite{seo2023garbage}.

\begin{figure}[tp]
    \centering
    \includegraphics[width=0.9\linewidth]{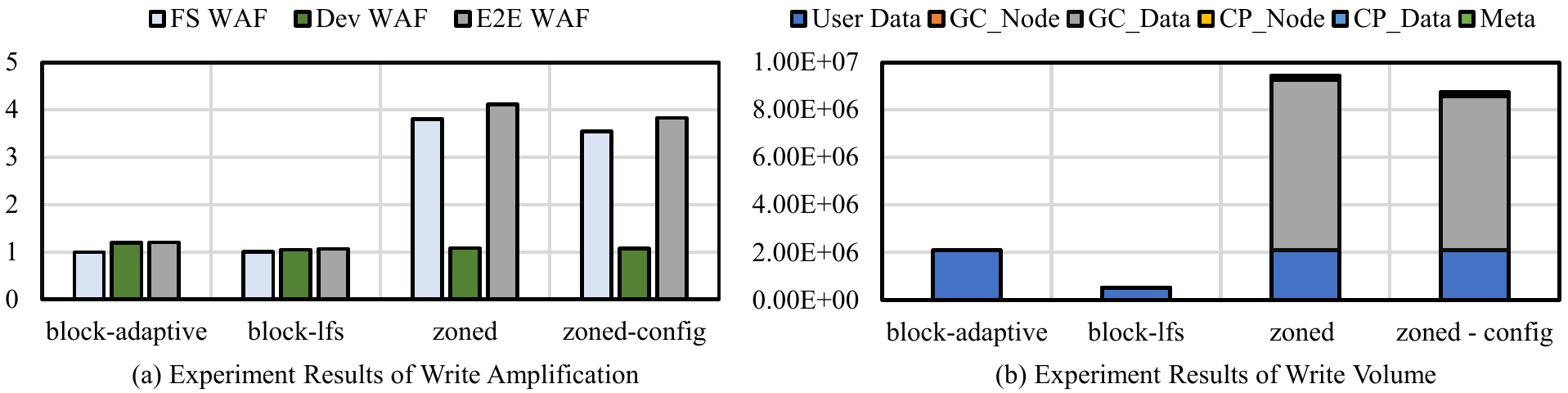}
    \caption{F2FS Garbage Collection and Write Volume Dynamics Across Zoned and Block Storage}
    \label{fig: casef2fs}
    \vspace{-0.1in}
\end{figure}

In the second experimental step, we observed a further widening of the bandwidth gap. 
Zoned storage exhibited a write bandwidth of 35.6 MiB/s (9.125 K IOPS), whereas block-adaptive achieved 84.9 MiB/s (21.7 K IOPS). 
This disparity is not only due to the larger write volume in zoned storage but also stems from an overly aggressive GC strategy.
As shown in Fig. \ref{fig: casef2fs}(a), the file system write amplification factor (FS WAF) for zoned storage was 3.81, while block-adaptive was very close to 1.
Fig. \ref{fig: casef2fs}(b) further illustrates that the file system write volume in zoned storage significantly exceeded that of block-adaptive, primarily driven by substantial GC writes.
Furthermore, we found that block-lfs could not complete the experiment smoothly due to insufficient file system space. 
From Fig. \ref{fig: casef2fs}(b), it is evident that the user data written by block-lfs was less than that of the other three comparison objects.
Moreover, even after adjusting the GC thresholds for F2FS on zoned storage, the situation did not significantly improve. 
The write bandwidth of zoned-config only increased to 38.1 MiB/s (9.766 K IOPS), and the file system write amplification decreased to 3.54. 
However, compared to block-adaptive, it still generated a large amount of GC, which interfered with user requests.
Despite the lower device write amplification (Dev WAF) of zoned storage, its end-to-end write amplification (E2E WAF) remained higher than that of block storage due to the excessively high file system write amplification. 
Through this experiment, we conclude that current file systems are still not adequately adapted for zoned storage. 
Future research and optimization in file system design are necessary to overcome the impact of the zoned interface's sequential write constraint on write performance and endurance.

\begin{figure}[tp]
    \centering
    \includegraphics[width=0.9\linewidth]{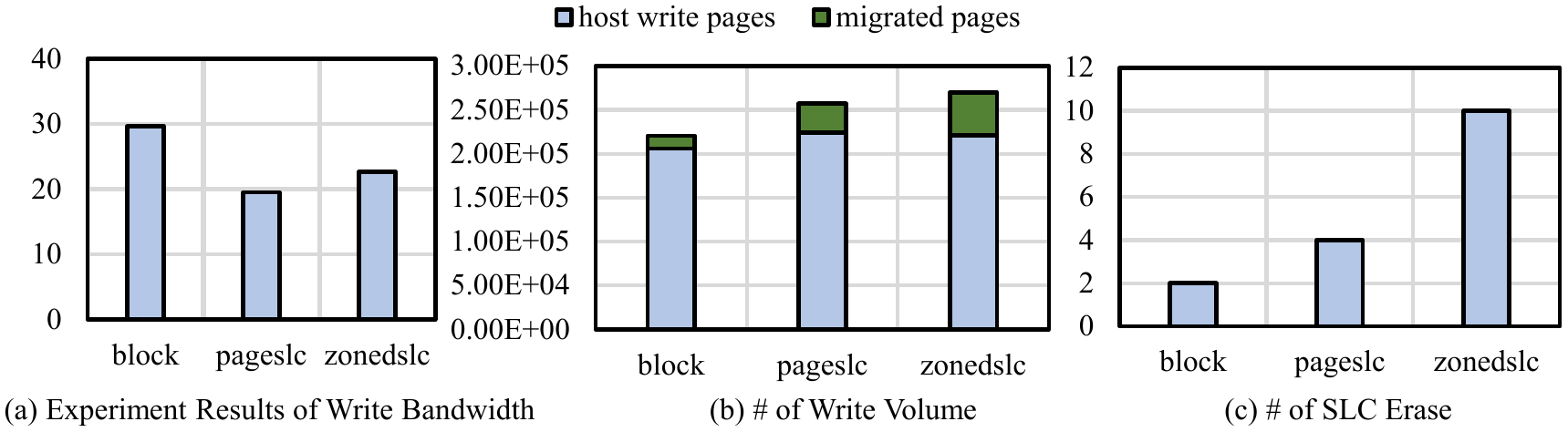}
    \caption{Impact of Data Migration Strategies in Zoned Storage}
    \label{fig: casemigrate}
    \vspace{-0.1in}
\end{figure}

\subsection{Case Study: Exploring Internal Data Migration Logic in Zoned Storage with F2FS}
When high-density flash memory like QLC is used, all user data is first written to the SLC layer (see Section \ref{subsec:bg_flash}). 
The management strategy for this SLC layer is critical in such scenarios. 
Ideally, short-lived data, which is erased quickly, should remain in SLC. 
This approach avoids writing it to QLC, thereby preventing unnecessary wear on the QLC. 
Consequently, we conduct this case study.
We use FIO to issue write streams with three distinct write hints: short, medium, and extreme. 
These hints inform F2FS about differences in data update frequency, leading to the data from these three streams being written into different zones.
Since using FIO with direct I/O and write hints causes errors, we employ buffered I/O, executing an \verb|fsync| after each I/O operation to simulate direct I/O behavior.
We configure the following experiment to simulate typical user behavior in consumer-grade scenarios\cite{zms}:
\begin{itemize}
    \item For short: We write 16 files, each 6 MiB, and then perform 8 random updates. 
    Following this, we select one file from all 16 to continue updating, totaling 160 MiB of updates.
    \item For medium: We write 16 files, each 12 MiB. 
    After performing 8 random updates, we introduce a 200~\textmu s sleep, then select one file from all 16 to continue updating, totaling 160 MiB of updates.
    \item For extreme: We write 32 files, each 4 MiB, with no subsequent updates.
\end{itemize}
The write granularity for all streams is 4 KiB.
For comparison, in addition to using block storage as a baseline, we also compare two different SLC management approaches within zoned storage: page-managed SLC (pageslc) and zone-managed SLC (zonedslc).
The experimental results, shown in Fig. \ref{fig: casemigrate}, indicate that block storage achieves a higher bandwidth of 29.7 MiB/s compared to zoned storage. 
This is because F2FS on block storage performs fewer writes and allows in-place updates, resulting in less data migration. 
As depicted in Fig. \ref{fig: casemigrate}(b), when SLC is managed by zones, the migration granularity increases, leading to a higher migration volume for zonedslc compared to pageslc.
However, Fig. \ref{fig: casemigrate}(a) shows that zonedslc achieves a bandwidth of 22.7 MiB/s, which is 16\% higher than pageslc. 
This improvement occurs because zonedslc can absorb more updates from short-lived data within the SLC. 
Fig. \ref{fig: casemigrate}(c) illustrates that zonedslc exhibits more direct SLC erasures than both pageslc and block storage.
Through this experiment, we conclude that future work can further optimize zoned storage's write performance by passing temperature information to it.

\subsection{Case Study: Read Performance with Different L2P Search Strategy}\label{case}
\begin{figure}[htp]
    \centering
    \includegraphics[width=0.6\linewidth]{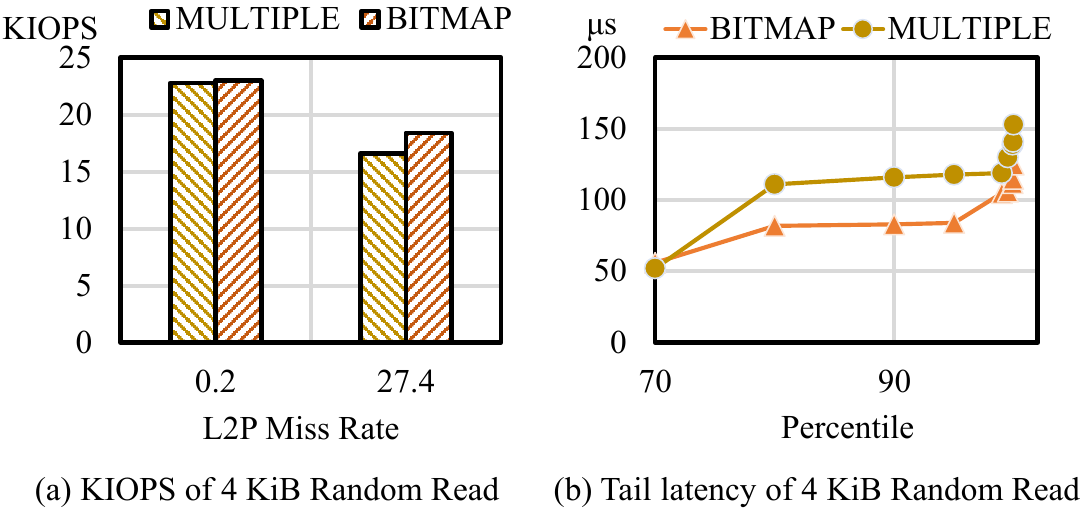}
    \caption{Impact of L2P Search Strategy on Random Reads with Hybrid Map}
    \label{fig: caseread}
    \vspace{-0.2in}
\end{figure}

The cost of an L2P cache miss is higher in hybrid mapping due to the multiple fetches required for L2P mapping entries. 
Fig. \ref{fig: caseread} compares the impact of L2P misses when using performance-optimized (BITMAP) and capacity-optimized (MULTIPLE) strategies.
Specifically, BITMAP uses a single bitmap, allowing the controller to instantly determine the mapping granularity for a given LPA without multiple page table lookups. 
When the L2P miss rate reaches 27.4\%, the KIOPS of MULTIPLE is 10\% lower than BITMAP, and its tail latency is also higher.
One feasible solution to mitigate this is to pin aggregated L2P mapping entries in the L2P cache once they are generated. 
When an L2P mapping entry with a larger mapping range is created, the previously covered L2P mapping entries are evicted. 
In the hybrid mapping mechanism, all mapping entries can be aggregated into a single zone mapping entry once a zone is full. 
Assuming a zone size of 16 MiB and an L2P cache entry size of 4 B, only 256 KiB of volatile memory is needed to cache all entries for 1 TiB of flash storage.
This capacity overhead is acceptable, and this solution is implemented as a configurable option in \nameplus.

\section{Conclusion}\label{sec:conclusion}
This paper designs a simulation platform that accounts for the unique read, write, and erase path designs of consumer-grade zoned storage. 
To further enhance the usability of this simulation platform, we integrate extensions for the block storage interface to support file system metadata updates. 
Additionally, we incorporate support for per-chip request queues, flexible block management, and compatibility with block and zone sizes that are not powers of two.
Finally, this paper validates the accuracy of the proposed \nameplus~platform through experiments and further explores the benefits and challenges inherent in zoned storage.
Moreover, we conduct three case studies, individually exploring the adaptability of zoned storage with F2FS, the management strategies for SLC flash blocks in zoned storage, and the L2P mapping table query strategies for zoned storage, providing valuable references for future research.

\bibliographystyle{ACM-Reference-Format}
\bibliography{ref}

\end{document}